\documentclass[conference]{IEEEtran}
\IEEEoverridecommandlockouts
\usepackage{graphicx}
\usepackage{subcaption} % Provides the subfigure environment
\usepackage{multirow} % Required for \multirow in Table 9
\usepackage{graphicx}  % Required for \resizebox{\textwidth}{!}
\usepackage{booktabs}  % Optional, but recommended for clean tables
\usepackage{cite}
\usepackage{amsmath,amssymb,amsfonts}
\usepackage{algorithmic}
\usepackage{graphicx}
\usepackage{textcomp}
\usepackage{xcolor}
\def\BibTeX{{\rm B\kern-.05em{\sc i\kern-.025em b}\kern-.08em
    T\kern-.1667em\lower.7ex\hbox{E}\kern-.125emX}}

\usepackage{tabularx}
\begin{document}

\title{Hierarchical Personalized Federated Learning for Fine-Grained Intrusion Detection: A Multi-Class Analysis in Heterogeneous Attack Environments}

\author{\IEEEauthorblockN{1\textsuperscript{st} Abul Kalam Azad}
\IEEEauthorblockA{\textit{Department of Computer Science} \\
\textit{The University of Alabama}\\
Tuscaloosa, AL, USA  \\
aazad1@crimson.ua.edu}
\and
\IEEEauthorblockN{2\textsuperscript{nd} Travis Atkison}
\IEEEauthorblockA{\textit{Department of Computer Science} \\
\textit{The University of Alabama}\\
Tuscaloosa, AL, USA \\
atkison@cs.ua.edu}
}
\maketitle
\begin{abstract}
Intrusion Detection Systems (IDS) play a vital role in protecting modern network infrastructures. Although Federated Learning (FL) enables collaborative training without sharing raw network traffic, global models struggle with non-IID attack distributions. Moreover, practical security operations require fine-grained attack classification for targeted threat responses. To address these challenges, this paper investigates the convergence dynamics and performance of Hierarchical Personalized Federated Learning (HPFL), which combines multi-tier aggregation with client-level personalization, for 8-class, 6-class, and 4-class intrusion detection using Multi-Layer Perceptron (MLP) models implemented in NVIDIA FLARE on the CICIDS2017 dataset. Experimental results show that personalized HPFL models achieve micro F1-scores up to 0.9911 in the 6-class task, while the overall local accuracy reaches 0.9975 in the 8-class task. On 8-class local test sets, HPFL achieves a micro F1-score of 0.9901 and a macro F1-score of 0.8324, outperforming FedAvg (0.7176) and FedProx (0.5624) in macro F1. Cross-client evaluations demonstrate out-of-distribution generalization, with HPFL achieving a 4-class micro F1-score of 0.9294 compared with 0.9140 for standalone local models. Furthermore, distance-weighted network analysis estimates a 48.9\% reduction in transmission overhead compared with conventional FL (194.4 MB vs. 380.3 MB for the 8-class task). These findings demonstrate HPFL's potential as an accurate, privacy-preserving, and communication-efficient approach to fine-grained intrusion detection in heterogeneous networks.
\end{abstract}


\section{Introduction}
\label{sec:introduction}
The rapid adoption of 5G/6G networks~\cite{oukebdane2025unraveling}, the Internet of Things (IoT)~\cite{choudhary2024internet}, wireless sensor networks~\cite{azad2019dcds}, cloud computing, and edge computing~\cite{andriulo2024edge} has increased network connectivity and data generation while expanding the attack surface. Intrusion Detection Systems (IDS) are therefore essential to protect these infrastructures. Although Machine Learning (ML)-based IDSs can effectively detect malicious traffic, conventional approaches often require centralized data collection, raising privacy and security concerns and conflicting with organizational policies.

Federated Learning (FL) addresses these concerns by enabling collaborative training without sharing raw data. However, global FL models struggle with non-IID network traffic arising from variations in attack signatures and benign distributions across clients. Hierarchical Personalized Federated Learning (HPFL) combines multi-tier aggregation with client-level adaptation to address this heterogeneity. Nevertheless, existing HPFL approaches primarily focus on binary classification~\cite{azad2026hierarchical}, leaving fine-grained multiclass intrusion detection underexplored. Extending HPFL to this setting presents three challenges: (1) severe class imbalance and non-IID distributions can hinder learning of rare or site-specific attacks; (2) changes in attack-label granularity can alter class distributions and affect convergence; and (3) stronger personalization can compromise cross-client generalization.

These challenges motivate a systematic investigation of HPFL for fine-grained multiclass intrusion detection. Accordingly, this study formulates three research questions: \textbf{RQ1}: Can HPFL effectively learn minority attack classes under imbalanced and non-IID distributions? \textbf{RQ2}: How do different label granularities affect convergence and personalization–generalization performance? \textbf{RQ3}: Can hierarchical personalization improve cross-client generalization while maintaining communication efficiency compared with conventional and personalized FL approaches?

To investigate these questions, the study evaluates HPFL under severe multiclass label skew using Multi-Layer Perceptron (MLP) models implemented in NVFlare~\cite{roth2022nvidia} on the CICIDS2017 dataset~\cite{sharafaldin2018toward}. The main contributions are as follows:

\begin{itemize}
    \item \textbf{Fine-Grained HPFL Analysis:} Characterizes multi-tier personalized aggregation under highly imbalanced attack distributions, showing its effectiveness in learning minority attack classes while ensuring global model stability.
    \item \textbf{Granularity Trade-Offs:} Systematically evaluates 8-class, 6-class, and 4-class classification to characterize the effects of label granularity on convergence and personalization/generalization performance.
    \item \textbf{Cross-Client Generalization Analysis:} Evaluates global, regional, and personalized models across client-specific and external client test distributions to characterize the generalization behavior of hierarchical personalization.
    \item \textbf{Comprehensive Benchmarking:}    
    Compares HPFL against centralized learning, isolated local models, FedAvg, FedProx, and FedRep, including an analysis of communication and computational costs.
\end{itemize}

\section{Related Work}
\label{sec:related_work}
ML techniques have been widely adopted for IDS. Conventional approaches, such as the hybrid CNN-KNN-SVM framework~\cite{gamal2020hybrid} and the Transformer-CNN model~\cite{kamal2024advanced}, achieve high detection accuracy on benchmark datasets. However, these centralized architectures require aggregating sensitive network logs into a single repository, creating privacy concerns in distributed environments.

FL addresses these privacy issues by enabling collaborative model training without raw data exchange. Early implementations, such as FELIDS~\cite{friha2022felids} presented multi-class IDS for IoT networks using standard FL. Similarly, Fed-ANIDS~\cite{idrissi2023fed} combined autoencoder-based anomaly detection with FL, while LSTM-based FL-IDS~\cite{anwar2025federated} improved detection accuracy over centralized approaches. Subsequent efforts focused on operational efficiency; for instance, lightweight CNN architectures with dynamic ensemble aggregation~\cite{baich2025federated}, threat intelligence sharing across heterogeneous networks~\cite{sarhan2023cyber}, and structured model pruning in Lightweight-Fed-NIDS~\cite{bouayad2024lightweight} significantly reduced resource overhead. Nevertheless, these frameworks rely on standard FedAvg~\cite{mcmahan2017communication} principles with a single aggregation tier, limiting their ability to handle severe statistical heterogeneity (non-IID data) across distributed clients.

To mitigate statistical heterogeneity and improve client-level performance, Personalized FL (PFL) has emerged as a promising paradigm. Existing PFL techniques generally fall into distinct categories: \textbf{Global Model Fine-Tuning:} These approaches learn a shared global representation and personalize it locally. For example, FedRep~\cite{collins2021exploiting} decouples model parameters into a shared global feature representation and client-specific heads, enabling local fine-tuning on client-specific data distributions. \textbf{Regularized \& Dual Model Personalization:} Other algorithms train dedicated local models constrained by global parameters. FedProx~\cite{li2020federated} adds a proximal term to regularize local updates against the global model, preventing divergence under non-IID conditions.

While PFL strategies such as FedProx and FedRep enhance local adaptation, they often encounter scalability issues when deployed across large networks. Hierarchical FL (HFL) addresses these issues by introducing multi-tier aggregation structures. HFed-IDS~\cite{sun2022hierarchical} and FedHSQA~\cite{xing2025fedhsqa} utilize intermediate servers to reduce communication overhead and improve aggregation robustness in smart grids and vehicular networks. However, standard HFL prioritizes communication efficiency, and lacks support for local attack heterogeneity. The HPFL framework ~\cite{azad2026hierarchical} focuses exclusively on binary classification and lacks comprehensive baseline comparisons. However, scaling multi-tier aggregation to fine-grained multi-class settings introduces severe tier-wise label skew. 

To address this limitation, this paper performs a rigorous diagnostic analysis of HPFL under multi-class conditions across 8-class, 6-class, and 4-class granularities. It analyzes personalization–generalization trade-offs and benchmarks performance against centralized bounds, global FL (FedAvg, FedProx), personalized FL (FedRep), and isolated local models.

\section{HPFL Framework for Fine-Grained Multi-Class Intrusion Detection}
\label{sec:framework}

\subsection{Architecture Overview}
\label{subsec:III-A}
\begin{figure}[!t]
    \centering
    \includegraphics[width=\linewidth]{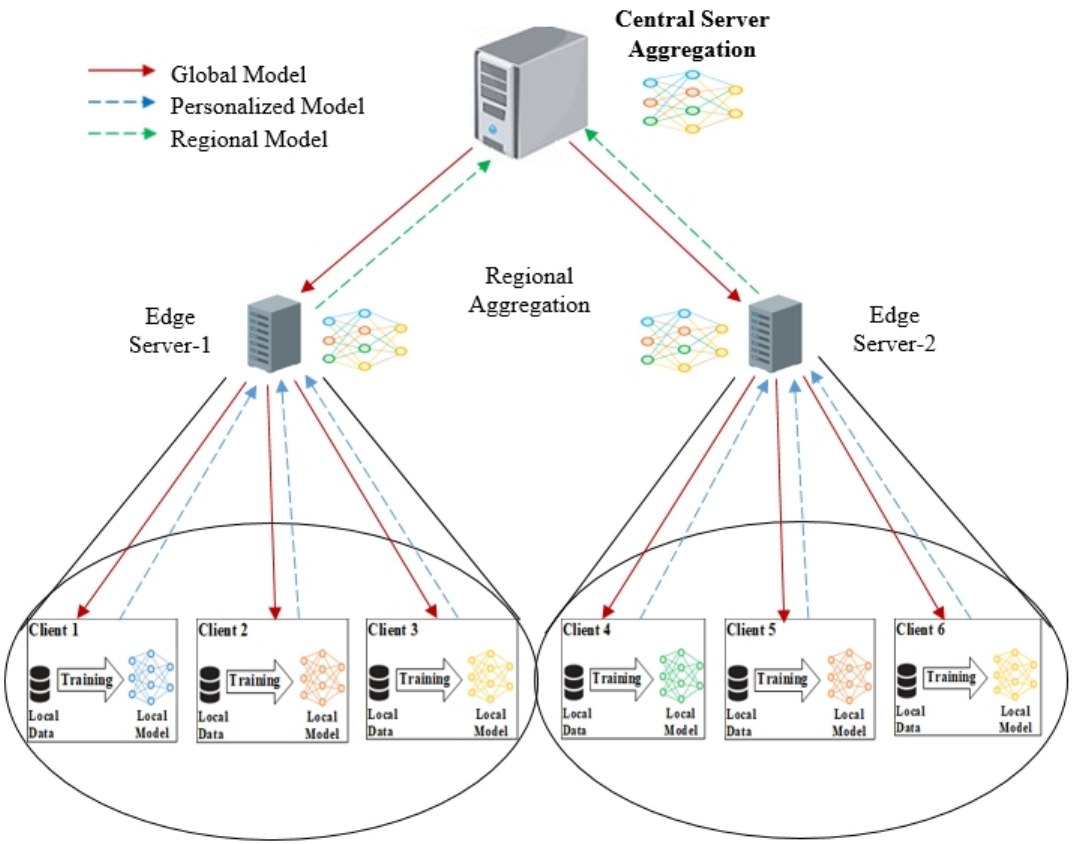}
    \caption{Overview of the proposed multi-class HPFL architecture for intrusion detection.}
    \label{fig:hpfl_architecture}
\end{figure}

The HPFL framework consists of a central server, edge servers, and multiple clients. The architecture enables collaborative model training while preserving data privacy and supporting adaptation to heterogeneous attack environments. The central server coordinates global model aggregation, edge servers perform regional aggregation, and clients perform personalized model training using their local data. As shown in Figure~\ref{fig:hpfl_architecture}, global model distributions follow the solid red lines, regional aggregations follow the dashed green lines, and personalized model updates are exchanged along the dashed blue lines between edge servers and local clients.

\subsection{Hierarchical Personalized Learning Process}
\label{subsec:III-B}

The training workflow operates across three sequential phases to account for severe class imbalances and regional traffic variations:

\subsubsection{Global Knowledge Learning}
\label{subsubsec:global_learning}
During the initial $m$ communication rounds (HFL rounds), client model updates are aggregated hierarchically through regional edge servers up to the central server. This builds a generalized global base model $W_{\text{global}}$, evaluated against a central test set to select the optimal global model parameters.

\subsubsection{Regional Adaptation}
\label{subsubsec:regional_adaptation}
The global parameters $W_{\text{global}}$ are distributed to regional edge servers for $n$ adaptation rounds. Edge servers aggregate parameters strictly within their assigned client clusters, tuning the model to regional attack frequencies while isolating unrelated network noise.

\subsubsection{Local Personalization}
\label{subsubsec:local_personalization}
Each client node initializes its classifier using the optimized regional weights and fine-tunes parameters on its local training dataset. This generates the final HPFL personalized model $W_{\text{pers}}^{i}$ for client $i$, optimized directly for its specific local attack distribution.

\subsection{Fine-Grained Attack Classification Strategy}
\label{subsec:III-C}
To analyze the impact of label granularity under non-IID conditions, the multi-class detection task is structured into three distinct configurations: 8-class, 6-class, and 4-class setups. The 8-class setup evaluates fine-grained recognition across distinct threat vectors, whereas the 6-class and 4-class setups group lower-frequency categories to measure how reduced label granularity influences model convergence and personalization performance. Detailed mappings appear in Section~\ref{subsec:IV-B}.

\section{Experimental Design}
\label{sec:experimental_design}

\subsection{Dataset Description and Preprocessing}
\label{subsec:IV-A}

\textbf{CICIDS2017 Dataset:} For these experiments, the publicly available CICIDS2017 benchmark dataset~\cite{sharafaldin2018toward} was utilized. The raw dataset contains 2,830,743 network traffic samples, each represented by 79 features collected over 5 days across benign conditions and 14 distinct attack categories.

\textbf{Data Cleaning and Feature Preparation:} For dataset cleaning, duplicate records were removed first, resulting in 2,499,736 remaining samples. Next, records containing invalid values (\textit{NaN} or infinite values) were removed. Features with constant values across all samples were also eliminated because they provide no discriminative signal. Afterward, a feature-correlation analysis was performed to identify redundant attributes; eight such features were detected and removed. Following these steps, the final cleaned dataset comprised 2,499,736 samples represented by 63 features. 

\subsection{Multi-Class Attack Grouping and Hierarchical Non-IID Partitioning}
\label{subsec:IV-B}
\textbf{Multi-Class Attack Grouping:} To test how class granularity affects performance, the study grouped benign traffic and the 14 CICIDS2017 attack types into three classification setups based on threat behavior:

\textbf{\textit{8-Class Setup:}} Fine-grained categories that keep distinct attack types separate (e.g., DoS, DDoS, and Web Attacks).

\textbf{\textit{6-Class Setup:}} Medium-grained categories that group attacks by functional goal (e.g., Flooding and Covert Exploits).

\textbf{\textit{4-Class Setup:}} Coarse-grained categories that merge attacks into high-level threat objectives (e.g., Flooding and Access Attacks). Table~\ref{tab:attack_grouping} shows the complete mapping for each setup.

% Table- I
\begin{table}[b]
\caption{Attack Grouping Strategy for the 8-Class, 6-Class, and 4-Class Classification Tasks}
\label{tab:attack_grouping}
\centering
\footnotesize % Compact font footprint for double-column layout
\setlength{\tabcolsep}{2pt} % Optimizes padding between columns
\begin{tabularx}{\columnwidth}{|X|l|l|l|}
\hline
\textbf{Original Attack Labels} & \textbf{8-Class} & \textbf{6-Class} & \textbf{4-Class} \\ \hline
BENIGN & BENIGN & BENIGN & BENIGN \\ \hline
DoS Hulk, DoS GoldenEye, DoS slowloris, DoS Slowhttptest & DoS & Flooding & Flooding \\ \hline
DDoS & DDoS & Flooding & Flooding \\ \hline
PortScan & PortScan & Reconnaissance & Reconnaissance \\ \hline
FTP-Patator, SSH-Patator & Brute Force & Credential Attack & Access Attack \\ \hline
Web Attack-Brute Force, Web Attack-XSS, Web Attack-SQL Injection & Web Attack & Web Attack & Access Attack \\ \hline
Bot & Botnet & Covert Exploit & Access Attack \\ \hline
Infiltration, Heartbleed & Other & Covert Exploit & Access Attack \\ \hline
\end{tabularx}
\end{table}

\textbf{Hierarchical Non-IID Data Partitioning:} To simulate realistic and severe threat heterogeneity, this study construct a multi-tier non-IID environment comprising a global server, two intermediate sites (A and B), and three clients per site ($C_1, C_2, C_3$). Specifically, 1\% of the dataset is reserved for global model initialization ($\ge 1$ sample/class) and 5\% for global evaluation. The remaining data is partitioned across Sites A and B via class-skewed allocation, assigning 95\% of site-dominant attack classes to their primary site to mimic localized threat exposure. Each site sets aside 10\% for site-level testing, distributing the remaining 90\% among its clients according to the task-specific specialization ratios detailed in Table~\ref{tab:non_iid_distribution}. Across all scenarios, benign traffic is distributed across clients at a 40\%/30\%/30\% ratio. Finally, client data is split into local train, validation, and test sets (70\%/10\%/20\%). Training and validation sets are downsampled to 30\% to simulate localized data scarcity, while local test sets remain intact for unbiased evaluation. All input features undergo standard z-score normalization.

% Table II
\begin{table}[b]
\caption{Hierarchical Non-IID Data Distribution Across Sites and Clients}
\label{tab:non_iid_distribution}
\centering
\scriptsize
\setlength{\tabcolsep}{2pt}
\renewcommand{\arraystretch}{1.15}
% Define exact widths totaling 1.0\columnwidth (accounting for borders)
\begin{tabular}{|p{0.1\columnwidth}|p{0.16\columnwidth}|p{0.16\columnwidth}|p{0.50\columnwidth}|}
\hline
\textbf{Task} & \textbf{Site A (95\%)} & \textbf{Site B (95\%)} & \textbf{Client Split ($C_1/C_2/C_3$)} \\ \hline
\textbf{8-Class} & DoS, DDoS, Botnet & PortScan, Patator, Web & \textbf{A:} DoS (90/5/5\%), DDoS (5/90/5\%), Botnet (5/5/90\%)\newline \textbf{B:} PortScan (90/5/5\%), Patator (5/90/5\%), Web (5/5/90\%) \\ \hline
\textbf{6-Class} & Flooding, Recon. & Credential, Web, Covert & \textbf{A:} Flooding (45/45/10\%), Recon (5/5/90\%)\newline \textbf{B:} Cred (90/5/5\%), Web (5/90/5\%), Covert (5/5/90\%) \\ \hline
\textbf{4-Class} & Flooding & Access, Recon. & \textbf{A:} Flooding (40/30/30\%)\newline \textbf{B:} Access (45/45/10\%), Recon (5/5/90\%) \\ \hline
\end{tabular}
\end{table}

\subsection{Implementation Details and Baseline Models}
\label{subsec:IV-C}

\textbf{System Configuration:} All experiments were conducted using NVFlare 2.4.1 in Python on an Ubuntu 20.04.6 LTS workstation equipped with an Intel Core i9-7900X CPU (3.30~GHz), an NVIDIA GeForce RTX 2080 Ti GPU, and 128~GB of RAM.

\textbf{HPFL Implementation Details:} Local clients train an MLP network with ReLU activations, outputting class logits $z_{j,c}$ for the target granularity ($C \in \{8, 6, 4\}$). Logits are converted to probabilities $\hat{y}_{j,c}$ via Softmax:
\begin{equation}
\label{eq:softmax}
\hat{y}_{j,c} = \frac{\exp(z_{j,c})}{\sum_{k=1}^{C} \exp(z_{j,k})}.
\end{equation}
Optimization uses the multi-class cross-entropy loss:
\begin{equation}
\label{eq:cross_entropy}
\mathcal{L}_{\text{CE}} = -\frac{1}{N} \sum_{j=1}^{N} \sum_{c=1}^{C} y_{j,c} \log(\hat{y}_{j,c}),
\end{equation}
where $N$ is the batch size and $y_{j,c}$ is the one-hot ground-truth label. Models are optimized using the Adam optimizer with weight decay. Specific hyperparameters across the 8-class, 6-class, and 4-class configurations are detailed in Section~\ref{subsec:IV-D}.

\textbf{Baseline Models:} Five representative architectures were evaluated across all classification tasks:
\begin{itemize}
\item \textbf{Centralized Model:} Trains a single classifier on pooled data, serving as a non-privacy-preserving empirical upper bound.
\item \textbf{Standard FedAvg~\cite{mcmahan2017communication}:} Learns a shared global model via iterative parameter averaging, omitting intermediate tiers and personalization.
\item \textbf{FedProx~\cite{li2020federated}:} Incorporates a proximal loss penalty to mitigate client drift under severe non-IID conditions.
\item \textbf{FedRep~\cite{collins2021exploiting}:} Decouples learning into a shared global feature extractor and locally fine-tuned classification heads.
\item \textbf{Local Models:} Trains isolated classifiers locally without cross-client collaboration, serving as a lower bound.
\end{itemize}
To ensure rigorous benchmarking, HPFL and all baselines share identical task-specific MLP backbones and hyperparameter configurations within each granularity (Section~\ref{subsec:IV-D}).

\subsection{Hyperparameter Tuning and Evaluation Metrics}
\label{subsec:IV-D}

\textbf{Hyperparameter Optimization:} This study used grid search to find the best model architectures and training settings for each task, selecting the configuration with the highest validation PR-AUC. Grid-search models were trained for 50 epochs using Xavier initialization without dropout. 

Table~\ref{tab:selected_hyperparameters} lists the selected parameters. To ensure a fair comparison, HPFL and all baselines used these exact same task-specific settings within each granularity. Specialized baselines were tuned using their own search spaces (e.g., FedProx $\mu$ tested between $0.0$ and $1.0$, and FedRep personal layers tested at $1$ and $2$). Finally, all experiments ran across three random seeds ($42$, $123$, and $999$), and reported the mean values.

% table III
\begin{table}[htbp]
\caption{Selected Hyperparameters and Training Configuration}
\label{tab:selected_hyperparameters}
\centering
\scriptsize
\setlength{\tabcolsep}{2.5pt}
\renewcommand{\arraystretch}{1.1}
\begin{tabularx}{\columnwidth}{|>{\raggedright\arraybackslash}X|c|c|c|}
\hline
\textbf{Parameter} & \textbf{8-Class} & \textbf{6-Class} & \textbf{4-Class} \\ \hline
HFL Rounds & 40 & 40 & 40 \\ \hline
PFL Rounds & 10 & 10 & 10 \\ \hline
Local Epochs/Round & 3 & 3 & 3 \\ \hline
Personalization Epochs & 10 & 10 & 10 \\ \hline
Model Selection Metric & PR-AUC & PR-AUC & PR-AUC \\ \hline
Loss Function & Cross Entropy & Cross Entropy & Cross Entropy \\ \hline
Input Features & 63 & 63 & 63 \\ \hline
Hidden Layers & [512, 256] & [512, 256, 128, 64] & [512, 256, 128, 64] \\ \hline
Activation & ReLU & ReLU & ReLU \\ \hline
Grid-Search Epochs & 50 & 50 & 50 \\ \hline
Optimizer & Adam & Adam & Adam \\ \hline
Learning Rate & 0.001 & 0.0001 & 0.001 \\ \hline
Weight Decay & $1 \times 10^{-5}$ & $1 \times 10^{-4}$ & $1 \times 10^{-5}$ \\ \hline
Batch Size & 64 & 128 & 128 \\ \hline
FedProx mu & 0.01 & 0.0001 & 0.0001 \\ \hline
FedRep Personal Layers & 1 & 1 & 1 \\ \hline
\end{tabularx}
\end{table}

\textbf{Evaluation Metrics:} To evaluate performance under extreme class imbalance, models were assessed using Accuracy, Precision, Recall, F1-score, FPR, FNR, and MCC. Because overall accuracy can be misleading on imbalanced data, F1-score and MCC serve as the primary metrics for minority threat detection. To account for class-wise and overall performance across the $K \in \{8, 6, 4\}$ classification tasks, this study utilized three averaging strategies.

\textbf{\textit{Macro Average:}} Computes the unweighted arithmetic mean across class-specific metrics $M_i$:
\begin{equation}
\label{eq:macro_avg}
M_{\text{macro}} = \frac{1}{K} \sum_{i=1}^{K} M_i.
\end{equation}
This gives equal importance to all classes, making it highly sensitive to minority attack performance.

\textbf{\textit{Support-Weighted Average:}} Weights each class metric $M_i$ by its sample support $s_i$ relative to total samples $S = \sum_{i=1}^{K} s_i$:
\begin{equation}
\label{eq:weighted_avg}
M_{\text{weighted}} = \frac{1}{S} \sum_{i=1}^{K} s_i M_i.
\end{equation}
This reflects the overall dataset distribution, where performance is dominated by majority classes (e.g., BENIGN).

\textbf{\textit{Micro Average:}} Aggregates confusion matrix (CM) counts across all $K$ classes before computing metrics. For example, micro-precision is defined as:
\begin{equation}
\label{eq:micro_precision}
\text{Precision}_{\text{micro}} = \frac{\sum_{i=1}^{K} \text{TP}_i}{\sum_{i=1}^{K} (\text{TP}_i + \text{FP}_i)},
\end{equation}
with similar formulations for micro-recall and micro-F1 (which are mathematically equivalent to overall accuracy).

Finally, to avoid bias from clients lacking certain attack classes, client-level CM counts ($\text{TP}_{i,c}, \text{FP}_{i,c}, \text{TN}_{i,c}, \text{FN}_{i,c}$) were aggregated across clients for each class $i$ before computing macro, support-weighted, and micro metrics.

% Section V
\section{Results and Discussion}
\label{sec:results}
This section evaluates the proposed multi-class HPFL framework for 8/6/4-class tasks. All metrics are averaged across three independent seeds (42, 123, 999) and six clients. The evaluation covers convergence dynamics, class-wise personalized performance, granularity impacts, multi-tier generalization, baseline comparisons, and communication efficiency.

% Subsection V-A
\subsection{Convergence Analysis Across Multi-Class Granularities}
\label{subsec:V-A}
\begin{figure*}[t]
\centering
\begin{subfigure}[b]{0.32\textwidth}
    \centering
    \includegraphics[width=\textwidth]{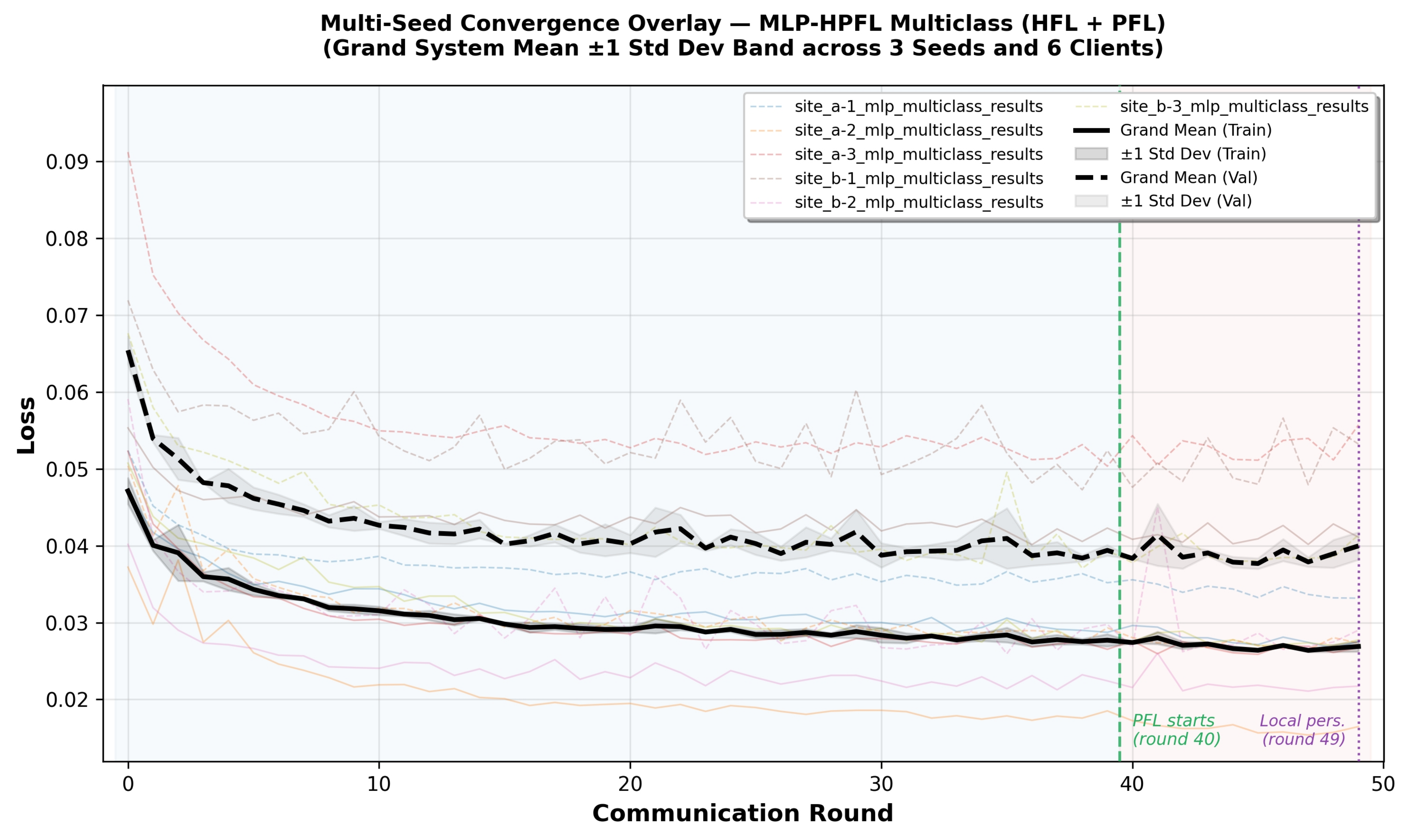}
    \caption{8-Class Task}
    \label{fig:2a}
\end{subfigure}\hfill
\begin{subfigure}[b]{0.32\textwidth}
    \centering
    \includegraphics[width=\textwidth]{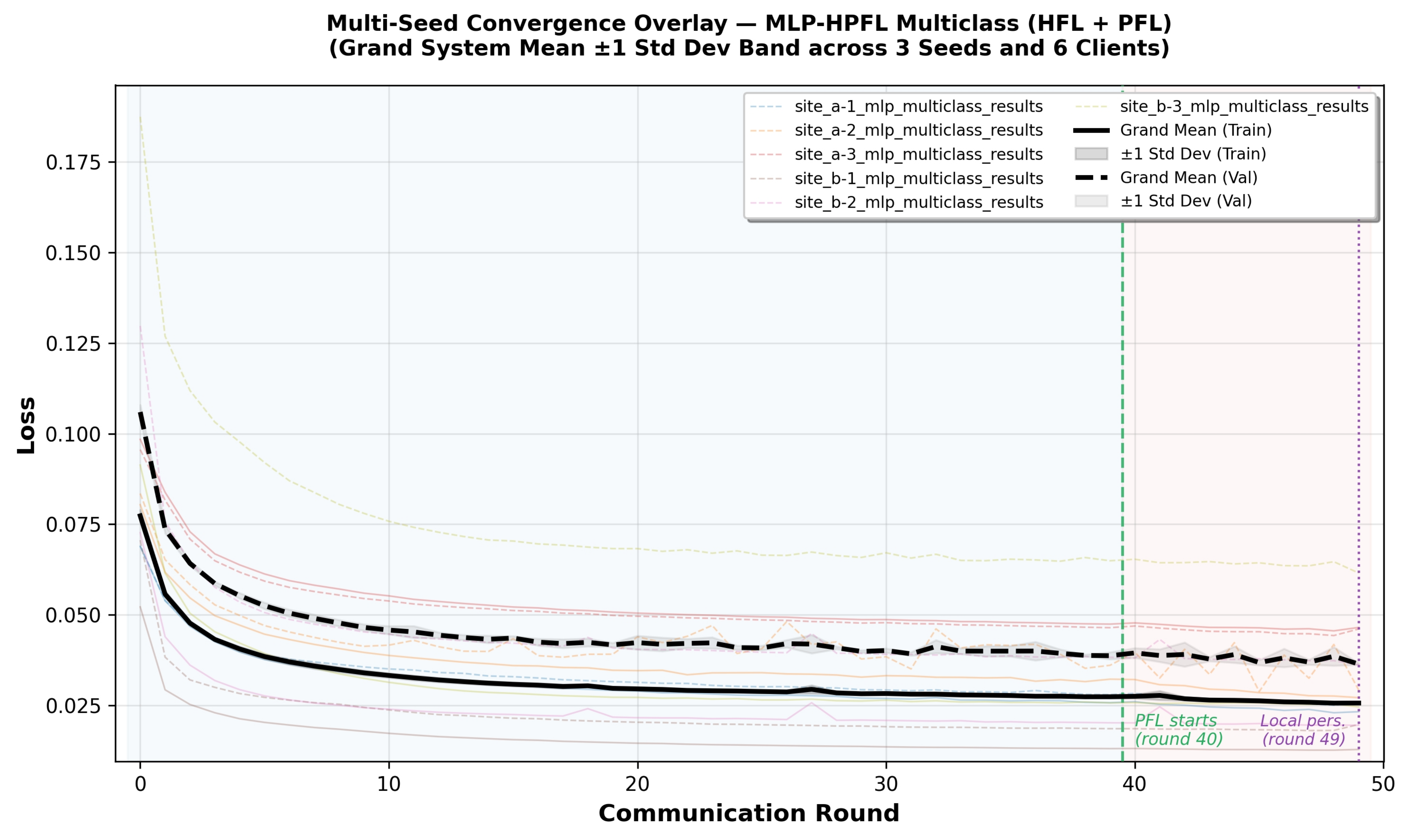}
    \caption{6-Class Task}
    \label{fig:2b}
\end{subfigure}\hfill
\begin{subfigure}[b]{0.32\textwidth}
    \centering
    \includegraphics[width=\textwidth]{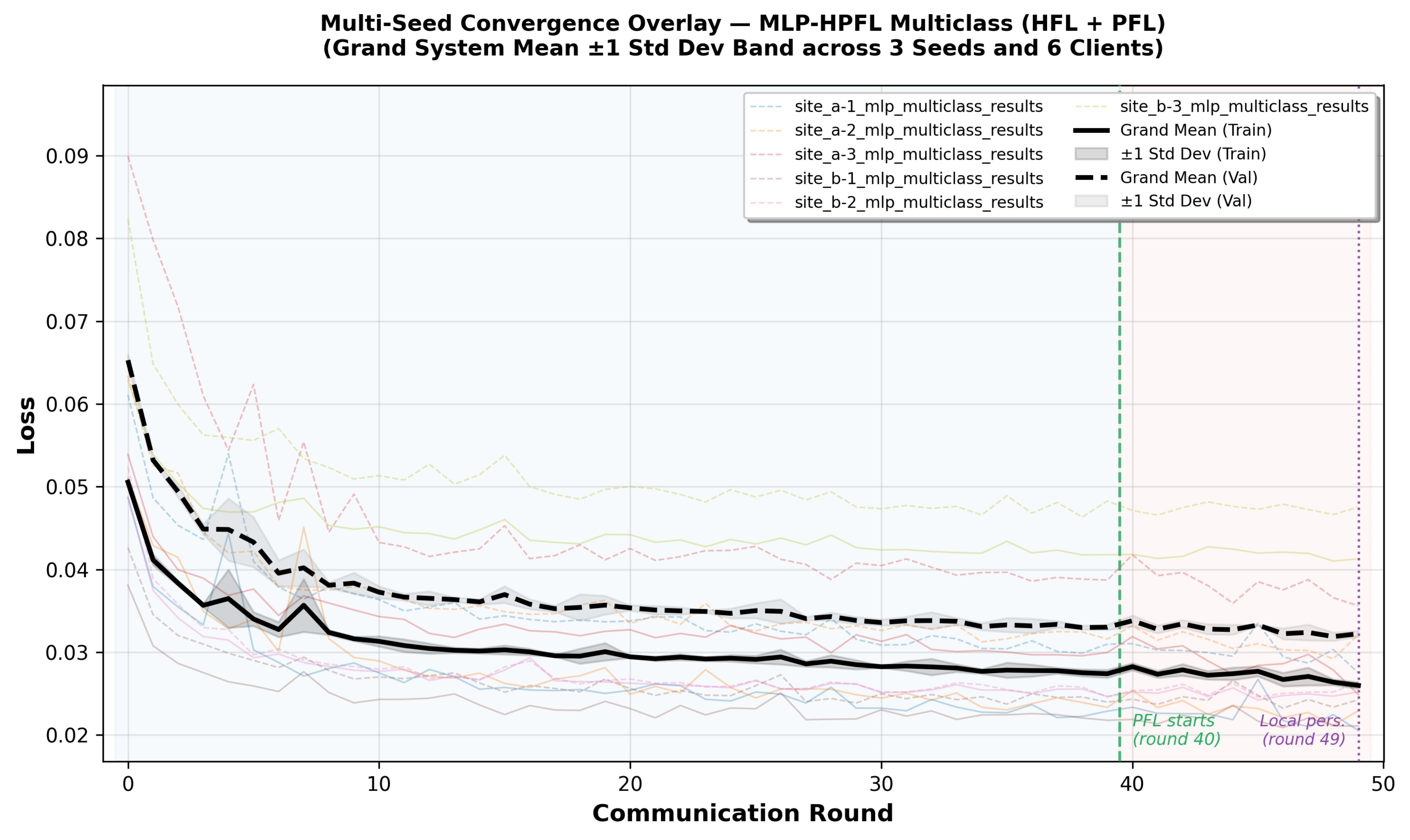}
    \caption{4-Class Task}
    \label{fig:2c}
\end{subfigure}
\caption{System-wide loss convergence trajectories and per-client learning dynamics for the MLP-HPFL model across (a) 8-class, (b) 6-class, and (c) 4-class. ($\pm 1$ std. dev.) across classification granularities: (a) 8-Class, (b) 6-Class, and (c) 4-Class.}
\label{fig:convergence_curves}
\end{figure*}

Figures~\ref{fig:2a}--\ref{fig:2c} illustrate the loss convergence trajectories over 50 communication rounds for the 8-class, 6-class, and 4-class MLP-HPFL setups, including $\pm1\sigma$ variance bands. Global learning (rounds 0--39) drives rapid loss decay within the first 10 rounds, with validation loss tracking closely above training loss without overfitting or divergence.

Transitioning to regional adaptation (round 40) and local personalization (round 49) maintains stability. Specifically, the 8-class setup (Fig.~\ref{fig:2a}) converges to ${\sim}0.027$ training and ${\sim}0.038$ validation loss despite severe non-IID heterogeneity. The 6-class task (Fig.~\ref{fig:2b}) yields a smooth decline (${\sim}0.026$ train, ${\sim}0.037$ val), while the 4-class setup (Fig.~\ref{fig:2c}) achieves the lowest error (${\sim}0.025$ train, ${\sim}0.032$ val). Overall, shifting from global aggregation to local personalization preserves stable optimization across all granularities.

% Figure 2 - Spans across two columns at top of page [t]

\subsection{Class-Wise and Average Performance of HPFL Personalized Models}
\label{subsec:V-B}

Tables~\ref{tab:hpfl_8class}, \ref{tab:hpfl_6class}, and \ref{tab:hpfl_4class} summarize the performance of personalized HPFL models across the 8-class, 6-class, and 4-class setups. Support-weighted and micro averages consistently exceed 0.9900 across all granularities, showing strong overall accuracy. However, macro averages reveal how class imbalance and rare sample sizes impact minority threat detection.

\subsubsection{8-Class Attack Classification}
As shown in Table~\ref{tab:hpfl_8class}, high-support classes achieve F1-scores for BENIGN (0.9941), DDoS (0.9965), and Brute Force (0.9788). However, performance drops on rare classes with low sample support, such as Botnet ($F_1 = 0.6690$), and Other ($F_1 = 0.4143$). This lowers the macro-average F1-score to $0.8324$, whereas support-weighted and micro F1 remain at $0.9901$.

% Table IV
\begin{table*}[t]
\centering
\caption{Class-Wise and Aggregate Performance of Personalized HPFL Model for 8-Class Classification on Local Test Data}
\label{tab:hpfl_8class}
\resizebox{\textwidth}{!}{%
\begin{tabular}{lcccccccc}
\hline
\textbf{Class} & \textbf{Accuracy} & \textbf{Precision} & \textbf{Recall} & \textbf{F1} & \textbf{FPR} & \textbf{FNR} & \textbf{MCC} & \textbf{Support} \\ \hline
Benign & 0.9902 & 0.9939 & 0.9942 & 0.9941 & 0.0295 & 0.0058 & 0.9652 & 350,666 \\
DoS & 0.9964 & 0.9817 & 0.9716 & 0.9765 & 0.0015 & 0.0284 & 0.9746 & 32,787 \\
DDoS & 0.9996 & 0.9984 & 0.9947 & 0.9965 & 0.0001 & 0.0053 & 0.9963 & 21,665 \\
PortScan & 0.9953 & 0.9267 & 0.9457 & 0.9361 & 0.0028 & 0.0543 & 0.9337 & 15,352 \\
Brute Force & 0.9998 & 0.9832 & 0.9745 & 0.9788 & 0.0001 & 0.0255 & 0.9787 & 1,553 \\
Web Attack & 0.9994 & 0.6004 & 0.8229 & 0.6943 & 0.0005 & 0.1771 & 0.7026 & 367 \\
Botnet & 0.9996 & 0.8401 & 0.5566 & 0.6690 & 0.0001 & 0.4434 & 0.6833 & 333 \\
Other & 1.0000 & 0.6556 & 0.3095 & 0.4143 & 0.0000 & 0.6905 & 0.4466 & 14 \\ \hline
Macro\_Avg & $0.9975 \pm 0.0003$ & $0.8725 \pm 0.0124$ & $0.8212 \pm 0.0167$ & $0.8324 \pm 0.0148$ & $0.0043 \pm 0.0010$ & $0.1788 \pm 0.0167$ & $0.8351 \pm 0.0143$ & -- \\
Support\_Weighted\_Avg & $0.9914 \pm 0.0010$ & $0.9903 \pm 0.0011$ & $0.9901 \pm 0.0011$ & $0.9901 \pm 0.0011$ & $0.0247 \pm 0.0067$ & $0.0099 \pm 0.0011$ & $0.9659 \pm 0.0039$ & -- \\
Micro\_Avg & $0.9901 \pm 0.0011$ & $0.9901 \pm 0.0011$ & $0.9901 \pm 0.0011$ & $0.9901 \pm 0.0011$ & $0.0014 \pm 0.0002$ & $0.0099 \pm 0.0011$ & $0.9887 \pm 0.0012$ & -- \\ \hline
\end{tabular}%
}
\end{table*}

\subsubsection{6-Class Attack Classification}
The 6-class task (Table~\ref{tab:hpfl_6class}) mitigates severe class imbalance by grouping minor attacks. Merging DoS and DDoS yields Flooding ($F_1 = 0.9852$), while consolidating Botnet, Infiltration, and Heartbleed into Covert Exploit raises its F1-score to $0.7065$. Consequently, the macro-average F1-score improves to $0.8845 \pm 0.0027$.

% Table V
\begin{table*}[t]
\centering
\caption{Class-Wise and Aggregate Performance of Personalized HPFL Model for 6-Class Classification on Local Test Data}
\label{tab:hpfl_6class}
\resizebox{\textwidth}{!}{%
\begin{tabular}{lcccccccc}
\hline
\textbf{Class} & \textbf{Accuracy} & \textbf{Precision} & \textbf{Recall} & \textbf{F1} & \textbf{FPR} & \textbf{FNR} & \textbf{MCC} & \textbf{Support} \\ \hline
Benign & 0.9912 & 0.9950 & 0.9943 & 0.9947 & 0.0241 & 0.0057 & 0.9688 & 350,664 \\
Flooding & 0.9962 & 0.9848 & 0.9857 & 0.9852 & 0.0023 & 0.0143 & 0.9830 & 54,447 \\
Reconnaissance & 0.9960 & 0.9387 & 0.9517 & 0.9452 & 0.0023 & 0.0483 & 0.9431 & 15,351 \\
Credential Attack & 0.9998 & 0.9601 & 0.9815 & 0.9707 & 0.0002 & 0.0185 & 0.9707 & 1,553 \\
Web Attack & 0.9994 & 0.6495 & 0.7711 & 0.7050 & 0.0004 & 0.2289 & 0.7074 & 367 \\
Covert Exploit & 0.9996 & 0.8726 & 0.5937 & 0.7065 & 0.0001 & 0.4063 & 0.7195 & 338 \\ \hline
Macro\_Avg & $0.9970 \pm 0.0002$ & $0.9001 \pm 0.0012$ & $0.8797 \pm 0.0024$ & $0.8845 \pm 0.0027$ & $0.0049 \pm 0.0004$ & $0.1203 \pm 0.0024$ & $0.8821 \pm 0.0025$ & -- \\
Support\_Weighted\_Avg & $0.9920 \pm 0.0006$ & $0.9911 \pm 0.0006$ & $0.9911 \pm 0.0006$ & $0.9911 \pm 0.0006$ & $0.0204 \pm 0.0019$ & $0.0089 \pm 0.0006$ & $0.9693 \pm 0.0021$ & -- \\
Micro\_Avg & $0.9911 \pm 0.0006$ & $0.9911 \pm 0.0006$ & $0.9911 \pm 0.0006$ & $0.9911 \pm 0.0006$ & $0.0018 \pm 0.0001$ & $0.0089 \pm 0.0006$ & $0.9893 \pm 0.0007$ & -- \\ \hline
\end{tabular}%
}
\end{table*}

\subsubsection{4-Class Attack Classification}
In the 4-class setup (Table~\ref{tab:hpfl_4class}), Flooding and Reconnaissance maintain robust F1-scores of $0.9827$ and $0.9444$, while access-related threats achieve an F1-score of $0.8631$. This coarse-grained grouping achieves the highest macro-average F1-score of $0.9461$ and macro MCC of $0.9378$, proving that reducing label granularity effectively counters non-IID data scarcity.

% Table VI
\begin{table*}[t]
\centering
\caption{Class-Wise and Aggregate Performance of Personalized HPFL Model for 4-Class Classification on Local Test Data}
\label{tab:hpfl_4class}
\resizebox{\textwidth}{!}{%
\begin{tabular}{lcccccccc}
\hline
\textbf{Class} & \textbf{Accuracy} & \textbf{Precision} & \textbf{Recall} & \textbf{F1} & \textbf{FPR} & \textbf{FNR} & \textbf{MCC} & \textbf{Support} \\ \hline
Benign & 0.9901 & 0.9949 & 0.9932 & 0.9940 & 0.0247 & 0.0068 & 0.9652 & 350,662 \\
Flooding & 0.9955 & 0.9783 & 0.9872 & 0.9827 & 0.0032 & 0.0128 & 0.9802 & 54,447 \\
Access Attacks & 0.9986 & 0.9030 & 0.8267 & 0.8631 & 0.0005 & 0.1733 & 0.8633 & 2,251 \\
Reconnaissance & 0.9959 & 0.9350 & 0.9539 & 0.9444 & 0.0025 & 0.0461 & 0.9423 & 15,352 \\ \hline
Macro\_Avg & $0.9950 \pm 0.0004$ & $0.9528 \pm 0.0052$ & $0.9403 \pm 0.0037$ & $0.9461 \pm 0.0008$ & $0.0077 \pm 0.0007$ & $0.0597 \pm 0.0037$ & $0.9378 \pm 0.0012$ & -- \\
Support\_Weighted\_Avg & $0.9911 \pm 0.0008$ & $0.9901 \pm 0.0008$ & $0.9901 \pm 0.0008$ & $0.9901 \pm 0.0008$ & $0.0210 \pm 0.0033$ & $0.0099 \pm 0.0008$ & $0.9657 \pm 0.0028$ & -- \\
Micro\_Avg & $0.9901 \pm 0.0008$ & $0.9901 \pm 0.0008$ & $0.9901 \pm 0.0008$ & $0.9901 \pm 0.0008$ & $0.0033 \pm 0.0003$ & $0.0099 \pm 0.0008$ & $0.9868 \pm 0.0011$ & -- \\ \hline
\end{tabular}%
}
\end{table*}

\subsection{Impact of Label Granularity}
\label{subsec:V-C}
\begin{figure}[t]
\centering
\begin{subfigure}[b]{0.48\linewidth}
    \centering
    \includegraphics[width=\linewidth]{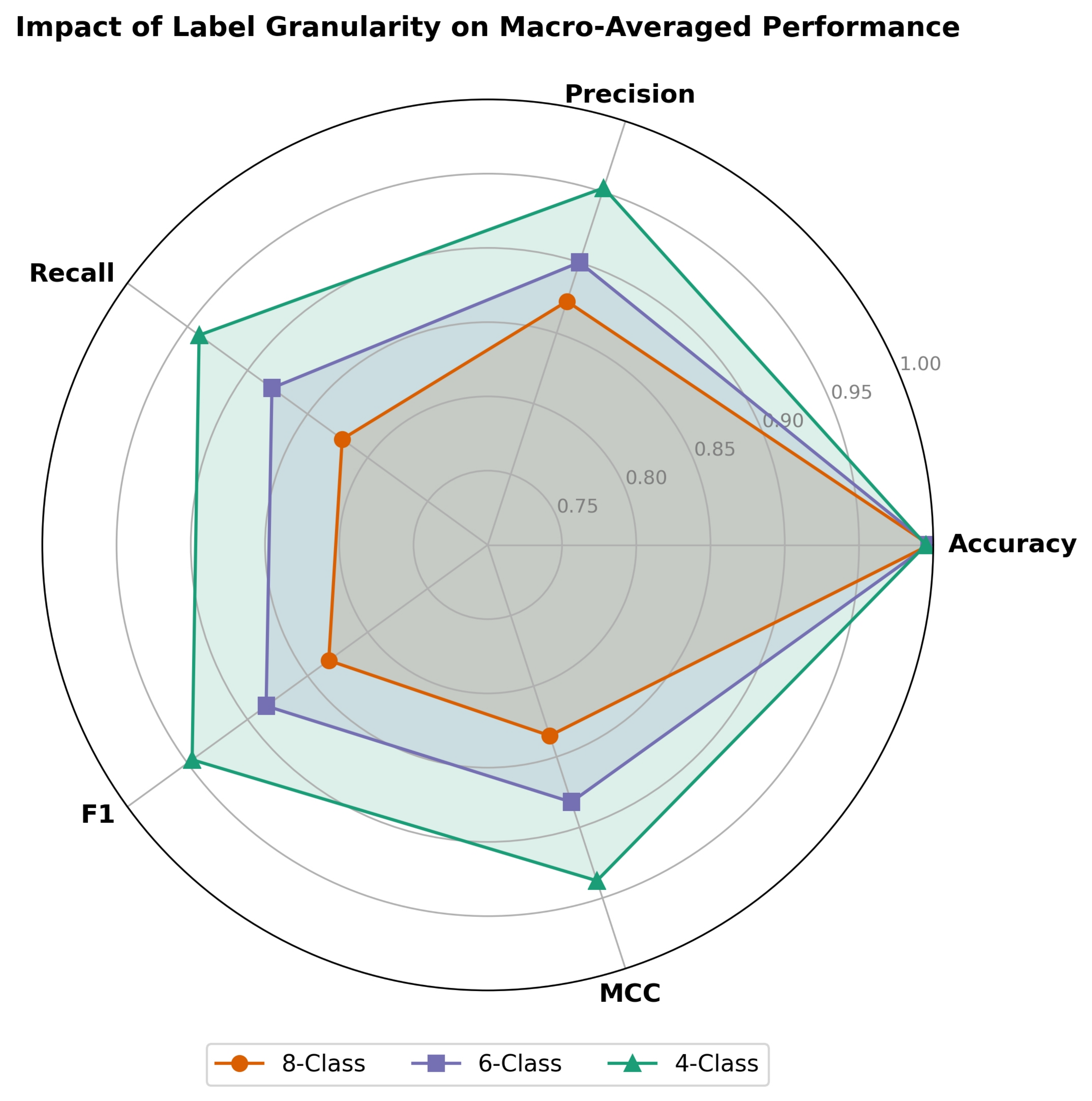}
    \caption{Macro-avg.}
    \label{fig:3a}
\end{subfigure}\hfill
\begin{subfigure}[b]{0.48\linewidth}
    \centering
    \includegraphics[width=\linewidth]{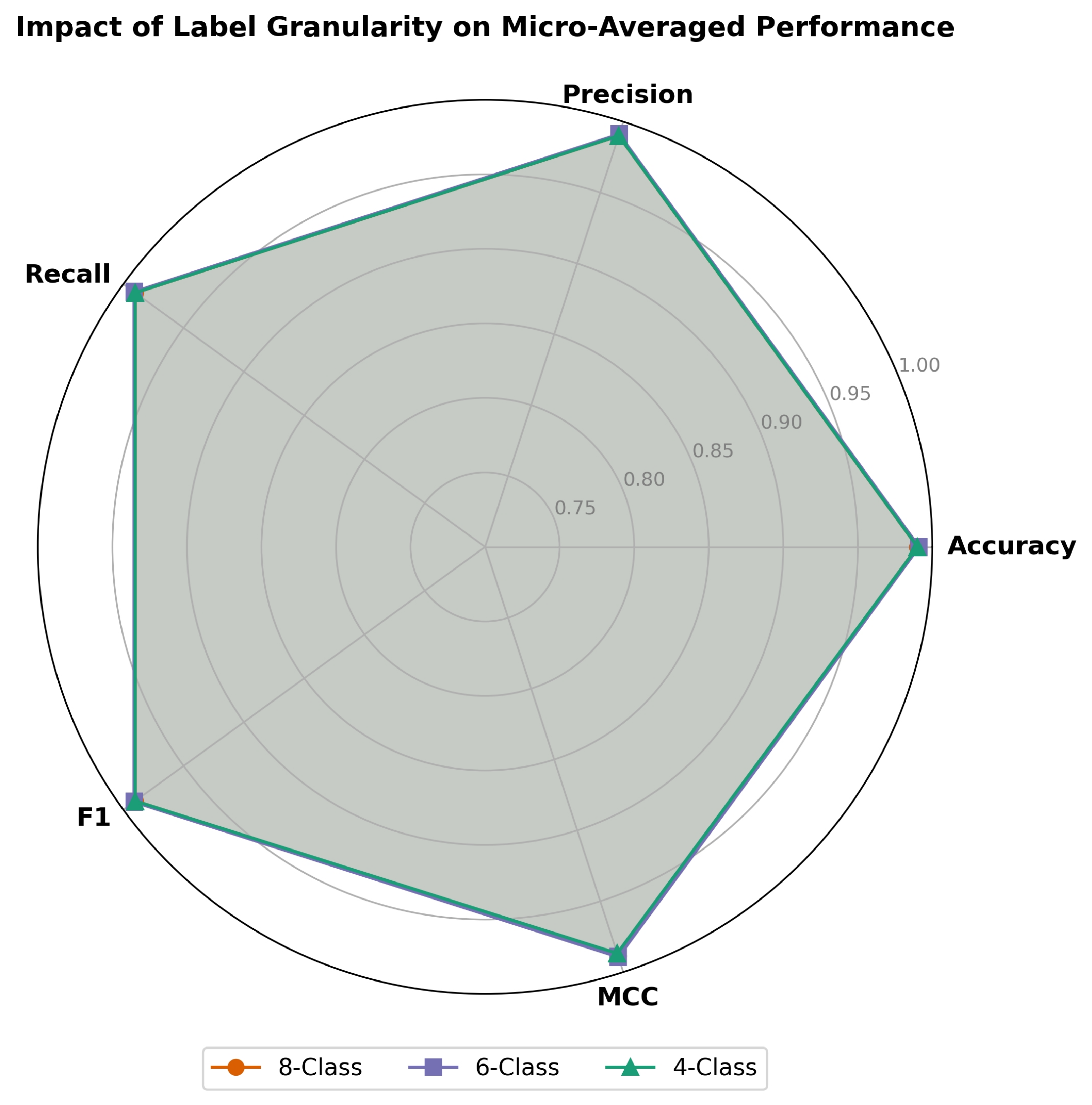}
    \caption{Micro-avg.}
    \label{fig:3b}
\end{subfigure}
\caption{Performance comparison of the personalized HPFL model across different class setups: (a) Macro-averaged, (b) Micro-averaged performance.}
\label{fig:radar_comparison}
\end{figure}

Figure~\ref{fig:radar_comparison} compares HPFL performance across the 8-class, 6-class, and 4-class setups using radar charts. As shown in Fig.~\ref{fig:3a}, macro-averaged scores improve as attack granularity decreases. Grouping rare attack types mitigates localized data scarcity, boosting macro Precision, Recall, F1, and MCC. In contrast, Fig.~\ref{fig:3b} shows that micro-averaged scores remain nearly constant ($>0.9900$) across all configurations.

% Figure 3 (Single Column, Side-by-Side Subfigures)

\subsection{Generalization Across HPFL Global, Regional, and Personalized Models}
\label{subsec:V-D}
Table~\ref{tab:hpfl_generalization} compares how HPFL global, regional, and personalized models perform across global, regional, and local test sets for 8-class task, demonstrating the structural benefits of the multi-tier HPFL framework. The global model achieves its highest accuracy on global test data (micro F1 of 0.9790 and MCC of 0.9760), but its performance drops on regional (0.9749) and local (0.9684) test sets because it struggles with non-IID data distributions. In contrast, the regional model peaks on regional test data with a micro F1-score of 0.9814 and an MCC of 0.9787, proving that intermediate edge aggregation successfully captures clustered network patterns. 

The personalized model shows the opposite trend, improving steadily as evaluation moves closer to the client node. Its micro F1-score rises from 0.9646 on global data to an optimal 0.9901 on local data (with a local micro MCC of 0.9887 and a peak macro F1-score of 0.8324). These findings confirm that the HPFL hierarchy operates precisely as intended: global learning shares broad features, regional aggregation models cluster characteristics, and local personalization delivers the highest defense accuracy for individual nodes.

% table VII
\begin{table*}[h]
\centering
\caption{Macro and Micro Average Performance of HPFL Global, Regional, and Personalized Models Across Global, Regional, and Local Test Environments for 8-Class Task}
\label{tab:hpfl_generalization}
\resizebox{\textwidth}{!}{%
\begin{tabular}{llccccc||ccccc}
\hline
\multirow{2}{*}{\textbf{Model Name}} & \multirow{2}{*}{\textbf{Test Data}} & \multicolumn{5}{c}{\textbf{Macro Average}} & \multicolumn{5}{c}{\textbf{Micro Average}} \\ \cline{3-12} 
 &  & \textbf{Accuracy} & \textbf{Precision} & \textbf{Recall} & \textbf{F1} & \textbf{MCC} & \textbf{Accuracy} & \textbf{Precision} & \textbf{Recall} & \textbf{F1} & \textbf{MCC} \\ \hline
\multirow{3}{*}{Global Model} & Global Test & 0.9948 & 0.8429 & 0.6444 & 0.6866 & 0.6990 & 0.9790 & 0.9790 & 0.9790 & 0.9790 & 0.9760 \\
 & Regional Test & 0.9937 & 0.8370 & 0.6766 & 0.7100 & 0.7155 & 0.9749 & 0.9749 & 0.9749 & 0.9749 & 0.9713 \\
 & Local Test & 0.9921 & 0.8460 & 0.6905 & 0.7189 & 0.7251 & 0.9684 & 0.9684 & 0.9684 & 0.9684 & 0.9639 \\ \hline
\multirow{3}{*}{Regional Model} & Global Test & 0.9938 & 0.8330 & 0.6318 & 0.6746 & 0.6844 & 0.9751 & 0.9751 & 0.9751 & 0.9751 & 0.9715 \\
 & Regional Test & 0.9953 & 0.8761 & 0.7025 & 0.7332 & 0.7405 & 0.9814 & 0.9814 & 0.9814 & 0.9814 & 0.9787 \\
 & Local Test & 0.9947 & 0.8621 & 0.6850 & 0.7231 & 0.7334 & 0.9788 & 0.9788 & 0.9788 & 0.9788 & 0.9758 \\ \hline
\multirow{3}{*}{Personalized Model} & Global Test & 0.9912 & 0.7824 & 0.5909 & 0.6388 & 0.6404 & 0.9646 & 0.9646 & 0.9646 & 0.9646 & 0.9596 \\
 & Regional Test & 0.9937 & 0.8020 & 0.6784 & 0.7223 & 0.7198 & 0.9749 & 0.9749 & 0.9749 & 0.9749 & 0.9714 \\
 & Local Test & 0.9975 & 0.8725 & 0.8212 & 0.8324 & 0.8351 & 0.9901 & 0.9901 & 0.9901 & 0.9901 & 0.9887 \\ \hline
\end{tabular}%
}
\end{table*}

\subsection{Comparative Analysis with Baseline Models}
\label{subsec:V-E}

Tables~\ref{tab:macro_baselines} and \ref{tab:micro_baselines} compare HPFL against centralized learning, standalone local training, and FL baselines (FedAvg, FedProx, FedRep) using macro and micro metrics.

\subsubsection{Macro-Average Personalization and Generalization}
As shown in Table~\ref{tab:macro_baselines}, HPFL consistently outperforms standard FL methods and matches or exceeds strong personalized baselines. In the 8-class task, HPFL achieves a personalized macro F1-score of 0.8324, outperforming FedAvg (0.7176), FedProx (0.5624), and FedRep (0.7892). On generalization, HPFL also surpasses FedRep and local training. These gains are largest for minority attack classes, where standard FL struggles with data heterogeneity. By using personalized local layers, HPFL adapts to local attack patterns to deliver higher Recall, F1, and MCC. In the 4-class task, HPFL reaches a macro F1 of 0.9461, closely matching the upper bound of centralized training (0.9507).

% Table VIII
\begin{table*}[t]
\centering
\caption{Macro-Average Personalization and Generalization Performance Comparison of HPFL and Baseline Models Across 8-Class, 6-Class, and 4-Class Classification Tasks}
\label{tab:macro_baselines}
\resizebox{\textwidth}{!}{%
\begin{tabular}{llccccc||ccccc}
\hline
\multirow{2}{*}{\textbf{Tasks}} & \multirow{2}{*}{\textbf{Models}} & \multicolumn{5}{c}{\textbf{Personalization}} & \multicolumn{5}{c}{\textbf{Generalization}} \\ \cline{3-12} 
 &  & \textbf{Accuracy} & \textbf{Precision} & \textbf{Recall} & \textbf{F1} & \textbf{MCC} & \textbf{Accuracy} & \textbf{Precision} & \textbf{Recall} & \textbf{F1} & \textbf{MCC} \\ \hline
\multirow{7}{*}{8-class} & Central Trained & 0.9978 & 0.8991 & 0.7856 & 0.7957 & 0.8106 & 0.9949 & 0.7863 & 0.6506 & 0.6840 & 0.6893 \\
 & FedAvg & 0.9918 & 0.8413 & 0.6914 & 0.7176 & 0.7222 & 0.9949 & 0.8253 & 0.6481 & 0.6867 & 0.6966 \\
 & FedProx & 0.9892 & 0.7467 & 0.5282 & 0.5624 & 0.5640 & 0.9932 & 0.7777 & 0.5368 & 0.5791 & 0.5896 \\
 & FedRep & 0.9976 & 0.8377 & 0.7763 & 0.7892 & 0.7926 & 0.9894 & 0.7168 & 0.5430 & 0.5809 & 0.5775 \\
 & Local Trained & 0.9977 & 0.8617 & 0.8259 & 0.8292 & 0.8322 & 0.9882 & 0.7453 & 0.5395 & 0.5905 & 0.5884 \\
 & \textbf{HPFL} & \textbf{0.9975} & \textbf{0.8725} & \textbf{0.8212} & \textbf{0.8324} & \textbf{0.8351} & \textbf{0.9912} & \textbf{0.7824} & \textbf{0.5909} & \textbf{0.6388} & \textbf{0.6404} \\ \hline
\multirow{7}{*}{6-class} & Central Trained & 0.9981 & 0.9424 & 0.7609 & 0.7792 & 0.8001 & 0.9953 & 0.8908 & 0.7197 & 0.7374 & 0.7449 \\
 & FedAvg & 0.9829 & 0.9329 & 0.6243 & 0.6772 & 0.6820 & 0.9926 & 0.9830 & 0.6616 & 0.7169 & 0.7289 \\
 & FedProx & 0.9830 & 0.9254 & 0.6225 & 0.6748 & 0.6792 & 0.9925 & 0.9831 & 0.6597 & 0.7155 & 0.7275 \\
 & FedRep & 0.9969 & 0.9024 & 0.8794 & 0.8850 & 0.8826 & 0.9844 & 0.7949 & 0.5974 & 0.6398 & 0.6252 \\
 & Local Trained & 0.9977 & 0.9114 & 0.8858 & 0.8933 & 0.8919 & 0.9817 & 0.6983 & 0.5250 & 0.5799 & 0.5607 \\
 & \textbf{HPFL} & \textbf{0.9970} & \textbf{0.9001} & \textbf{0.8797} & \textbf{0.8845} & \textbf{0.8821} & \textbf{0.9850} & \textbf{0.8213} & \textbf{0.6057} & \textbf{0.6509} & \textbf{0.6434} \\ \hline
\multirow{7}{*}{4-class} & Central Trained & 0.9958 & 0.9722 & 0.9333 & 0.9507 & 0.9444 & 0.9909 & 0.9738 & 0.8495 & 0.9026 & 0.8913 \\
 & FedAvg & 0.9865 & 0.9547 & 0.8619 & 0.9036 & 0.8822 & 0.9901 & 0.9758 & 0.8468 & 0.9021 & 0.8893 \\
 & FedProx & 0.9866 & 0.9596 & 0.8575 & 0.9029 & 0.8820 & 0.9898 & 0.9774 & 0.8405 & 0.8987 & 0.8861 \\
 & FedRep & 0.9807 & 0.6473 & 0.6336 & 0.6396 & 0.6182 & 0.9706 & 0.6234 & 0.5713 & 0.5909 & 0.5555 \\
 & Local Trained & 0.9957 & 0.9729 & 0.9289 & 0.9483 & 0.9421 & 0.9756 & 0.8733 & 0.7054 & 0.7571 & 0.7292 \\
 & \textbf{HPFL} & \textbf{0.9950} & \textbf{0.9528} & \textbf{0.9403} & \textbf{0.9461} & \textbf{0.9378} & \textbf{0.9803} & \textbf{0.9170} & \textbf{0.7943} & \textbf{0.8427} & \textbf{0.8146} \\ \hline
\end{tabular}%
}
\end{table*}

\subsubsection{Micro-Average Personalization and Generalization Performance}
Table~\ref{tab:micro_baselines} shows micro-average performance. Because majority classes dominate total sample counts, overall scores across all models remain high. However, HPFL maintains near-optimal performance while outperforming conventional FL. In the 8-class setup, HPFL achieves a micro F1-score of 0.9901 and an MCC of 0.9887 on local test data. Similar strong results appear in the 6-class ($F_1 = 0.9911$) and 4-class ($F_1 = 0.9901$) tasks, proving that HPFL effectively handles non-IID data distributions.

% table IX
\begin{table*}[t]
\centering
\caption{Micro-Average Personalization and Generalization Performance Comparison of HPFL and Baseline Models Across 8-Class, 6-Class, and 4-Class Classification Tasks}
\label{tab:micro_baselines}
\resizebox{\textwidth}{!}{%
\begin{tabular}{llccccc||ccccc}
\hline
\multirow{2}{*}{\textbf{Tasks}} & \multirow{2}{*}{\textbf{Models}} & \multicolumn{5}{c}{\textbf{Personalization}} & \multicolumn{5}{c}{\textbf{Generalization}} \\ \cline{3-12} 
 &  & \textbf{Accuracy} & \textbf{Precision} & \textbf{Recall} & \textbf{F1} & \textbf{MCC} & \textbf{Accuracy} & \textbf{Precision} & \textbf{Recall} & \textbf{F1} & \textbf{MCC} \\ \hline
\multirow{7}{*}{8-class} & Central Trained & 0.9912 & 0.9912 & 0.9912 & 0.9912 & 0.9900 & 0.9797 & 0.9797 & 0.9797 & 0.9797 & 0.9768 \\
 & FedAvg & 0.9673 & 0.9673 & 0.9673 & 0.9673 & 0.9627 & 0.9797 & 0.9797 & 0.9797 & 0.9797 & 0.9768 \\
 & FedProx & 0.9567 & 0.9567 & 0.9567 & 0.9567 & 0.9505 & 0.9728 & 0.9728 & 0.9728 & 0.9728 & 0.9689 \\
 & FedRep & 0.9904 & 0.9904 & 0.9904 & 0.9904 & 0.9890 & 0.9578 & 0.9578 & 0.9578 & 0.9578 & 0.9518 \\
 & Local Trained & 0.9977 & 0.9909 & 0.9909 & 0.9909 & 0.9896 & 0.9882 & 0.9527 & 0.9527 & 0.9527 & 0.9460 \\
 & \textbf{HPFL} & \textbf{0.9901} & \textbf{0.9901} & \textbf{0.9901} & \textbf{0.9901} & \textbf{0.9887} & \textbf{0.9646} & \textbf{0.9646} & \textbf{0.9646} & \textbf{0.9646} & \textbf{0.9596} \\ \hline
\multirow{7}{*}{6-class} & Central Trained & 0.9943 & 0.9943 & 0.9943 & 0.9943 & 0.9931 & 0.9861 & 0.9861 & 0.9861 & 0.9861 & 0.9833 \\
 & FedAvg & 0.9488 & 0.9488 & 0.9488 & 0.9488 & 0.9386 & 0.9779 & 0.9779 & 0.9779 & 0.9779 & 0.9734 \\
 & FedProx & 0.9491 & 0.9491 & 0.9491 & 0.9491 & 0.9389 & 0.9775 & 0.9775 & 0.9775 & 0.9775 & 0.9730 \\
 & FedRep & 0.9908 & 0.9908 & 0.9908 & 0.9908 & 0.9889 & 0.9533 & 0.9533 & 0.9533 & 0.9533 & 0.9440 \\
 & Local Trained & 0.9977 & 0.9931 & 0.9931 & 0.9931 & 0.9917 & 0.9817 & 0.9450 & 0.9450 & 0.9450 & 0.9340 \\
 & \textbf{HPFL} & \textbf{0.9911} & \textbf{0.9911} & \textbf{0.9911} & \textbf{0.9911} & \textbf{0.9893} & \textbf{0.9549} & \textbf{0.9549} & \textbf{0.9549} & \textbf{0.9549} & \textbf{0.9459} \\ \hline
\multirow{7}{*}{4-class} & Central Trained & 0.9916 & 0.9916 & 0.9916 & 0.9916 & 0.9888 & 0.9818 & 0.9818 & 0.9818 & 0.9818 & 0.9758 \\
 & FedAvg & 0.9730 & 0.9730 & 0.9730 & 0.9730 & 0.9639 & 0.9801 & 0.9801 & 0.9801 & 0.9801 & 0.9735 \\
 & FedProx & 0.9733 & 0.9733 & 0.9733 & 0.9733 & 0.9644 & 0.9797 & 0.9797 & 0.9797 & 0.9797 & 0.9729 \\
 & FedRep & 0.9614 & 0.9614 & 0.9614 & 0.9614 & 0.9486 & 0.9413 & 0.9413 & 0.9413 & 0.9413 & 0.9217 \\
 & Local Trained & 0.9957 & 0.9914 & 0.9914 & 0.9914 & 0.9885 & 0.9756 & 0.9514 & 0.9514 & 0.9514 & 0.9351 \\
 & \textbf{HPFL} & \textbf{0.9901} & \textbf{0.9901} & \textbf{0.9901} & \textbf{0.9901} & \textbf{0.9868} & \textbf{0.9605} & \textbf{0.9605} & \textbf{0.9605} & \textbf{0.9605} & \textbf{0.9474} \\ \hline
\end{tabular}%
}
\end{table*}

\subsubsection{Cross-Client Performance Comparison with Local Models}
Table~\ref{tab:cross_client} evaluates out-of-distribution generalization by testing each client model on the combined test data of all other clients. HPFL consistently outperforms standalone local models across all tasks. In the 8-class setup, HPFL raises the macro F1-score from 0.5277 to 0.5781 and the micro F1-score from 0.9115 to 0.9278. 

This advantage is highest in the 4-class setup, where HPFL reaches a macro F1-score of 0.7531 and an MCC of 0.6974 (compared to 0.6797 and 0.6145 for local models). While local models overfit to their own data, HPFL uses shared global and regional layers to learn transferable features, allowing personalized models to accurately detect external client traffic.

% Table X
\begin{table*}[t]
\centering
\caption{Cross-Client Performance Comparison Between HPFL Personalized Models and Standalone Local Models}
\label{tab:cross_client}
\resizebox{\textwidth}{!}{%
\begin{tabular}{llccccc||ccccc}
\hline
\multirow{2}{*}{\textbf{Tasks}} & \multirow{2}{*}{\textbf{Models}} & \multicolumn{5}{c}{\textbf{Macro-Average}} & \multicolumn{5}{c}{\textbf{Micro-Average}} \\ \cline{3-12} 
 &  & \textbf{Accuracy} & \textbf{Precision} & \textbf{Recall} & \textbf{F1} & \textbf{MCC} & \textbf{Accuracy} & \textbf{Precision} & \textbf{Recall} & \textbf{F1} & \textbf{MCC} \\ \hline
\multirow{2}{*}{8-class} & Local Trained & 0.9779 & 0.6037 & 0.4899 & 0.5277 & 0.4986 & 0.9115 & 0.9115 & 0.9115 & 0.9115 & 0.8988 \\
 & \textbf{HPFL} & \textbf{0.9819} & \textbf{0.6539} & \textbf{0.5314} & \textbf{0.5781} & \textbf{0.5549} & \textbf{0.9278} & \textbf{0.9278} & \textbf{0.9278} & \textbf{0.9278} & \textbf{0.9175} \\ \hline
\multirow{2}{*}{6-class} & Local Trained & 0.9645 & 0.5660 & 0.4745 & 0.4993 & 0.4449 & 0.8935 & 0.8935 & 0.8935 & 0.8935 & 0.8722 \\
 & \textbf{HPFL} & \textbf{0.9680} & \textbf{0.6388} & \textbf{0.5018} & \textbf{0.5478} & \textbf{0.5000} & \textbf{0.9040} & \textbf{0.9040} & \textbf{0.9040} & \textbf{0.9040} & \textbf{0.8848} \\ \hline
\multirow{2}{*}{4-class} & Local Trained & 0.9570 & 0.7720 & 0.6298 & 0.6797 & 0.6145 & 0.9140 & 0.9140 & 0.9140 & 0.9140 & 0.8853 \\
 & \textbf{HPFL} & \textbf{0.9647} & \textbf{0.8251} & \textbf{0.7086} & \textbf{0.7531} & \textbf{0.6974} & \textbf{0.9294} & \textbf{0.9294} & \textbf{0.9294} & \textbf{0.9294} & \textbf{0.9058} \\ \hline
\end{tabular}%
}
\end{table*}

\subsection{Communication and Computational Cost Analysis}
\label{subsec:V-G}
To evaluate communication efficiency, this study used a normalized distance model. Client-to-central links in standard FL serve as the baseline (weight = $1.0$), while client-to-site and site-to-central links are assigned distance weights of $0.33$ and $0.67$, respectively.

\begin{figure}[t]
\centering
\includegraphics[width=\linewidth]{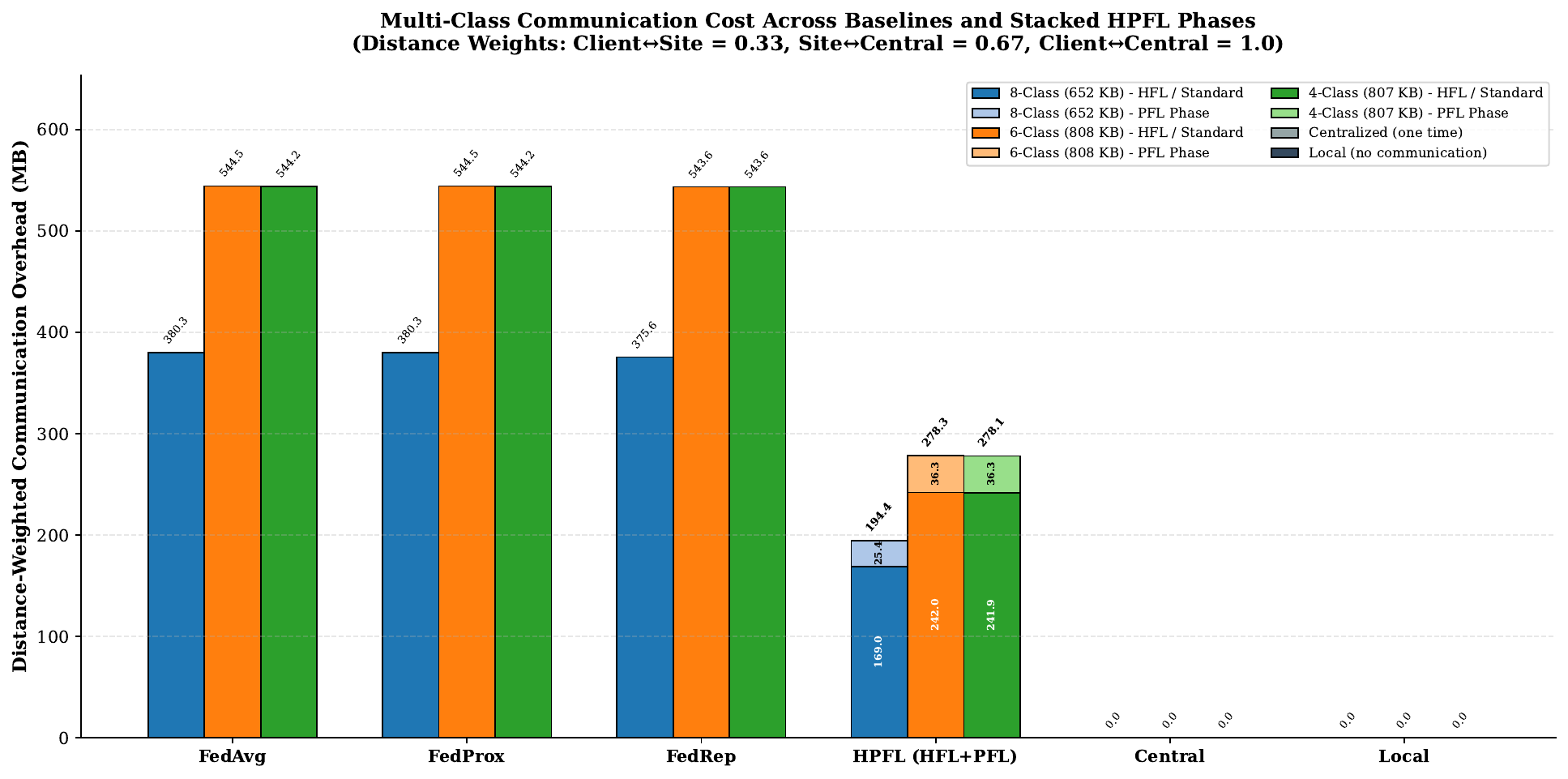}
\caption{Communication cost analysis among HPFL with Baseline Models.}
\label{fig:comm_cost}
\end{figure}

Figure~\ref{fig:comm_cost} compares the distance-weighted communication overhead of HPFL against FL baselines (FedAvg, FedProx, and FedRep). These FL approaches incur high overhead because all clients exchange parameters directly with the central server every round ($375.6\text{ MB}$ to $380.3\text{ MB}$ for 8-class; $543.6\text{ MB}$ to $544.5\text{ MB}$ for 6-class and 4-class). In contrast, HPFL routes client updates through intermediate site servers over low-cost links (weight = $0.33$). For the 8-class task, HPFL uses $169.0\text{ MB}$ in the HFL phase and $25.4\text{ MB}$ in the PFL phase ($194.4\text{ MB}$ total), achieving a $48.9\%$ cost reduction compared to FedAvg ($380.3\text{ MB}$). For 6-class and 4-class tasks, HPFL requires $278.3\text{ MB}$ and $278.1\text{ MB}$ total overhead, maintaining the same $48.9\%$ savings over the FL baselines ($544.2\text{ MB}$--$544.5\text{ MB}$). Centralized and local models require no cross-network parameter exchanges ($0\text{ MB}$).

% Figure 5 (Single Column)

HPFL also distributes computational workloads across tiers, with intermediate site servers processing client updates in parallel while the central server aggregates only site-level models and remains inactive during local personalization. This hierarchical design reduces central-server bottlenecks and supports scalable deployment.

\section{Conclusion}
\label{sec:conclusion}
This paper analyzed the systemic behavior, convergence dynamics, and generalization tradeoffs of HPFL for fine-grained multi-class intrusion detection. By evaluating HPFL across 8-class, 6-class, and 4-class scenarios on the CICIDS2017 dataset, the results demonstrated that multi-tier aggregation and client personalization effectively handle severe tier-wise label skew. Experimental results show that HPFL achieves micro F1-scores above 0.9900 and outperforms FL baselines, raising the 8-class macro F1-score to 0.8324 (compared to 0.7176 for FedAvg and 0.5624 for FedProx). Furthermore, cross-client evaluations confirm robust out-of-distribution generalization, while network modeling demonstrates a 48.9\% reduction in communication overhead. 

Future work will explore diverse architectures (such as CNNs and recurrent models) with additional datasets, such as CICIoT2023, dynamic client clustering, and differential privacy mechanisms. Overall, HPFL provides a scalable, communication-efficient, and accurate solution for fine-grained threat detection in heterogeneous networks including large-scale IoT, smart transportation, cloud computing, and enterprise network environments.

% Set the IEEE transaction bibliography style
\bibliographystyle{IEEEtran}

% Point to your references.bib file (do not include the .bib extension here)
\bibliography{references}

\end{document}